\documentclass[sigconf]{acmart}
\usepackage{hyperref}
\usepackage{url}
\usepackage{multirow}
\usepackage{cleveref}
\usepackage{float}
\usepackage{balance}
\usepackage{microtype}
\usepackage{color}
\usepackage{graphics,graphicx}
\usepackage{epstopdf}
\usepackage{colortbl}
\usepackage{soul}
\usepackage{subcaption}
\usepackage{enumitem}
\usepackage[justification=centering]{caption}

\AtBeginDocument{%
  \providecommand\BibTeX{{%
    \normalfont B\kern-0.5em{\scshape i\kern-0.25em b}\kern-0.8em\TeX}}}

\setcopyright{none}
\acmDOI{10.1145/3627508.3638335}

\begin{document}

\title{The Influence of Presentation and Performance on User Satisfaction}

\author{Kanaad Pathak}
\email{kanaad.pathak@strath.ac.uk}
\orcid{0000-0002-1246-8685}
\affiliation{%
  \institution{University of Strathclyde}
  \streetaddress{16 Richmond St}
  \city{Glasgow}
  \country{UK}
  \postcode{G1 1XQ}
}
\author{Leif Azzopardi}

\email{leif.azzopardi@strath.ac.uk}
\orcid{0000-0002-6900-0557}
\affiliation{%
  \institution{University of Strathclyde}
  \streetaddress{16 Richmond St}
  \city{Glasgow}
  \country{UK}
}
\author{Martin Halvey}
\email{martin.halvey@strath.ac.uk}
\orcid{0000-0001-6387-8679}
\affiliation{%
  \institution{University of Strathclyde}
  \streetaddress{16 Richmond St}
  \city{Glasgow}
  \country{UK}
  \postcode{G1 1XQ}
}
%

\renewcommand{\shortauthors}{Pathak, et al.}

\begin{abstract}

Information Retrieval (IR) systems are designed to provide users with a ranked list of results based on their queries. The effectiveness of an IR system is gauged not just by its ability to retrieve relevant results but also by how it presents these results to users; an engaging presentation often correlates with increased user satisfaction. While existing research has delved into the link between user satisfaction, IR performance metrics, and presentation, these aspects have typically been investigated in isolation. Our research aims to bridge this gap by examining the relationship between query performance, presentation and user satisfaction.
For our analysis, we conducted a between-subjects experiment comparing the effectiveness of various result card layouts for an ad-hoc news search interface. Drawing data from the TREC WaPo 2018 collection, we centered our study on four specific topics. Within each of these topics, we assessed six distinct queries with varying nDCG values. Our study involved 164 participants who were exposed to one of five distinct layouts containing result cards, such as ``title'', ``title+image'', or ``title+image+summary''.
Our findings indicate that while nDCG is a strong predictor of user satisfaction at the query level, there exists no linear relationship between the performance of the query, presentation of results and user satisfaction. However, when considering the total gain on the initial result page, we observed that presentation does play a significant role in user satisfaction (at the query level) for certain layouts with result cards such as, title+image or title+image+summary. Our results also suggest that the layout differences have complex and multifaceted impacts on satisfaction.
We demonstrate the capacity to equalize user satisfaction levels between queries of varying performance by changing how results are presented.
This emphasizes the necessity to harmonize both performance and presentation in IR systems, considering users' diverse preferences. Ultimately, our insights can steer the evolution of more user-aligned IR systems, underscoring the balance between system performance and result presentation.

\end{abstract}

\ccsdesc[500]{Information systems~Users and interactive retrieval}
\ccsdesc[500]{Human-centered computing~Empirical studies in HCI}

\keywords{
Information Retrieval (IR),
User Satisfaction,
Interface Layouts,
Query Performance,
Search Result Presentation,
Empirical Study,
Human-Computer Interaction,
Retrieval Effectiveness,
Search Interfaces,
}


\maketitle
\section{Introduction}
Information Retrieval (IR) systems, such as web search engines, aim to help users efficiently locate relevant information within vast collections of documents in response to queries. A critical aspect of an IR system's effectiveness lies in its ability to fulfil a user's information needs by retrieving documents relevant to the user's query. Retrieved documents are generally presented to users on Search Engine Result Pages (SERPs), and each result can typically be represented by a result card. A good result card aims to help the user make more effective decisions about exploring a given document by presenting on it, information such as a title, image or summary of the web page. Previous works from \citet{Rele2005UsingInterfaces, Teevan2009VisualRevisitation, Kammerer2010HowInterface} and \citet{Joho2006AWeb} have studied how the presentation of these result cards affects user satisfaction. The broad consensus from these analyses is that incorporating visual elements like images, links, and text summaries can strongly influence user satisfaction and perceptions of relevance. 

However, it is not solely the presentation that drives user satisfaction. Users also spend their time creating queries so that the system may retrieve and present them with appropriate relevant documents for their information need. The performance of these queries is typically measured by system-side metrics such as Cumulative Gain (CG), Discounted Cumulative Gain (DCG), normalized Discounted Cumulative Gain (nDCG) etc. Work from \citet{Al-Maskari2007TheSatisfaction} has explored how these metrics affect user satisfaction, finding that there is a strong correlation in most query performance metrics.

During the interaction process, users can also perform other actions such as inspecting various result cards, saving the documents behind the cards etc. These actions come with inherent costs to perform them and further work such as \citet{Azzopardi2011TheRetrievalb} have developed formal models to estimate the costs in the interaction process (such as cost to query, examine cards etc). Given this formal framework, further research such as~\citet{Azzopardi2013HowBehavior,Morrison2000Within-personBenefits,Verma2017SearchMobile} have studied how costs such as the cost to query affect user satisfaction.

Given that the presentation of the result cards can also affect user satisfaction, it is unclear how changing the presentation can affect both the system side costs (query costs) and user side costs (user satisfaction). Take, for example, two result lists with slightly differing nDCGs for a given query, presented in the same result card type (all titles). Findings from \citet{Al-Maskari2007TheSatisfaction,Morrison2000Within-personBenefits,Verma2017SearchMobile} would suggest that spending longer examining results for a query with a higher nDCG will lead to more satisfaction. However, if we modify the presentation of the result list with a lower nDCG to be presented with, say all titles and images (TI), users may now spend more time examining results in this list due to their changed presentation and thus feel a similar amount of satisfaction as that obtained from a result list with a higher nDCG. 

Our study seeks to explore this interplay between query performance, presentation, and user satisfaction. To ground our analysis, we present findings from a crowd-sourced user study relating to an ad-hoc news search task. In our study, we examine five interface layouts with differing numbers of results per page with four distinct types of result cards. Utilizing topics, queries, and documents from the TREC WaPo 2018 corpus, participants were tasked with finding and marking relevant documents for two topics. We collected satisfaction ratings from users for each query, and subsequently for each result card layout (after they completed a topic). The primary research questions guiding our study are:

\begin{enumerate}[label=\textbf{(RQ\arabic*)}]
\item  How do the quality of search results (as measured by query performance) and the interface layout impact user satisfaction in information retrieval tasks?
\item  What are the effects of different interface layouts on user satisfaction as measured by overall satisfaction, the likability of the engine, productivity, and mental effort?
\end{enumerate}
\section{Background}
\label{background}
Several factors can impact user satisfaction with an information retrieval (IR) system. 
These factors can be thought of as costs and be measured system-side or user-side to determine the overall effectiveness of an IR system. 
The system side costs include standard IR effectiveness metrics such as precision, cumulative gain (CG), discounted cumulative gain (DCG) and normalized DCG (nDCG). Where, on the user side the costs can include the formulation of queries, the display of results, time to browse, user satisfaction etc. 
On the system side, studies such from \citet{Al-Maskari2010ARetrieval} and \citet{Al-Maskari2007TheSatisfaction} have examined how IR effectiveness measures such as precision, CG, DCG and nDCG affect user satisfaction, and found that CG and precision have a better correlation with user satisfaction as compared to nDCG.
The correlation with nDCG was weak due to having limited judgements. On the system side, experimentation has been conducted independent of the presentation of results. Studies on the user side have also considered the effect of presentation on user satisfaction, mainly along three dimensions. 
In the first dimension, studies such as \citet{Rele2005UsingInterfaces} and \citet{Kammerer2010HowInterface} tell us that there are two main layouts in which results can be optimally presented to maximize user satisfaction. The first is a single list, and the second one is a grid layout. 
However, it is worth noting that both of these studies have found conflicting results on which layout is better. 
The second dimension explores the relationship between different result card formats and user satisfaction. 
Work done by \citet{Teevan2009VisualRevisitation, Dziadosz2002DoResults, Joho2006AWeb, Tombros1998AdvantagesRetrieval, Bota2016PlayingWorkload} has explored standard cards which contain information such as the URL, text and images. 
The first two dimensions (layout and presentation format) come with an important trade-off on the space and utility of each result item, which is explored in the third dimension. 
For example, if results are presented as a list of ten blue links versus if we present them with an image and some summary text, the results will occupy different amounts of space on the screen, and then satisfaction to the user will largely depend on the visual appeal and informativeness of the result \cite{Kelly2015HowExperience, Maxwell2017AExperience,Roy2022UsersExperiences}. 
Models developed on economic search theory by \citet{Azzopardi2015AnBehavior} have been proposed which provide a framework to estimate the costs associated with user behaviour, and studies such as \citet{Verma2017SearchMobile, Jansen2000RealWeb} have measured how costs associated with search, such as the length of the query, the number of viewed documents and clicked snippets affect user satisfaction. 
However, both on the system side and the user side, these costs have been studied independently of each other. 
That is to say; user satisfaction has not been studied in the context of the presentation of results and also standard IR metrics such as nDCG. The interplay between IR performance and presentation is still not well understood. 
\vspace{-0.1cm}
\section{Methodology}
\label{experiment}

To explore how query performance and result card layouts influence user satisfaction, we conducted a between-subjects study using simulated ad-hoc search tasks \cite{Borlund1997TheSystems}. To position the information-seeking process within a structured context, participants were presented with a series of pre-picked queries, which were grouped into three categories based on their nDCG@10: low, medium, and high. Each category contained two queries. The task involved participants engaging in an exploratory search session, examining various queries and documents to find and pinpoint relevant examples within relevant documents related to the given topic. All documents were indexed and retrieved using our custom-built system, ensuring consistency across searches and presentation of results. The between-group variable in our study was defined by five distinct interface layouts. These layouts prominently featured cards consisting of titles, images, and summaries of news articles.
\subsection{Collection and System}
\label{subsection:collection_and_system}
\begin{figure*}
\centering
\includegraphics[width=\textwidth]{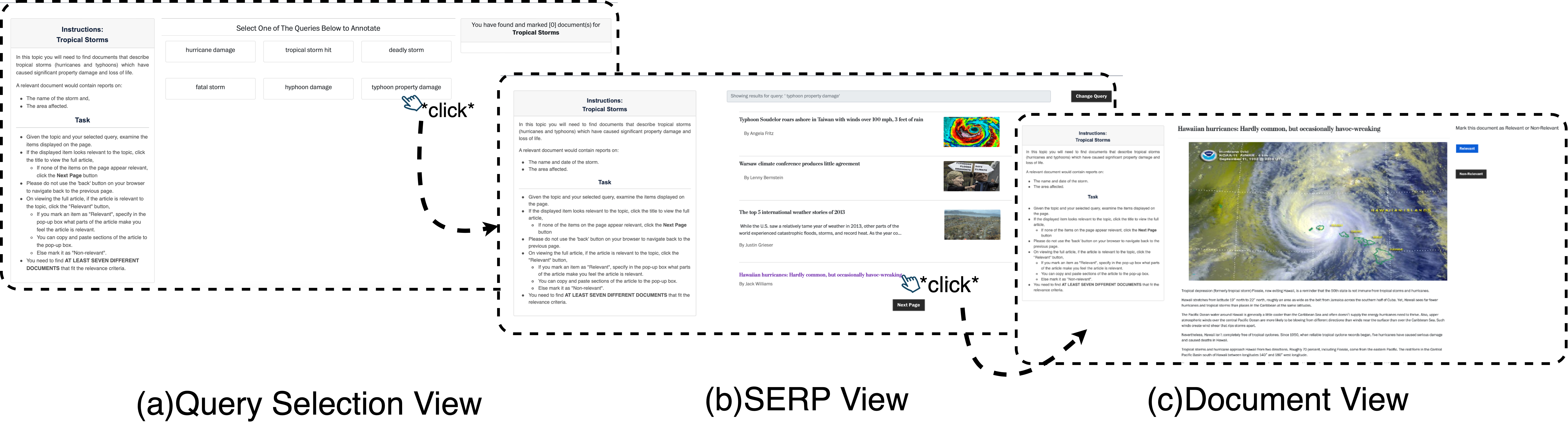}
\caption{An example of the user interface presented to participants for collection of annotations. Sub-figure (b) shows an example of a SERP layout with a random arrangement of cards.}
\label{fig:serp_study_ui}
\end{figure*}

For this study, we used the TREC Washington Post Corpus (WaPo) collection from the TREC Common Core 2018 track~\footnote{https://trec-core.github.io/2018/}. The WaPo collection consists of 608,180 news articles and blog posts published between January 2012 and August 2017 categorized into 50 topics for information retrieval tasks. This collection provides a diverse range of topics for analysis and experimentation, allowing us to explore the effectiveness of our proposed approach across different topical themes. We used \textit{\textbf{Whoosh}}\footnote{\$pip install whoosh==2.7.4} (a pure python search engine library) with BM25 ($b=0.75$, $k_1=1.2$) to index and retrieve documents for a given query. We presented results on a SERP, as shown in Figure\ref{fig:serp_study_ui}(b). Our SERP view consisted of result cards in presentation formats of two major news sources (The Washington Post and Google News).

\begin{figure*}[t]
     \centering
     \begin{subfigure}[b]{0.23\textwidth}
         \centering
         \includegraphics[width=\textwidth]{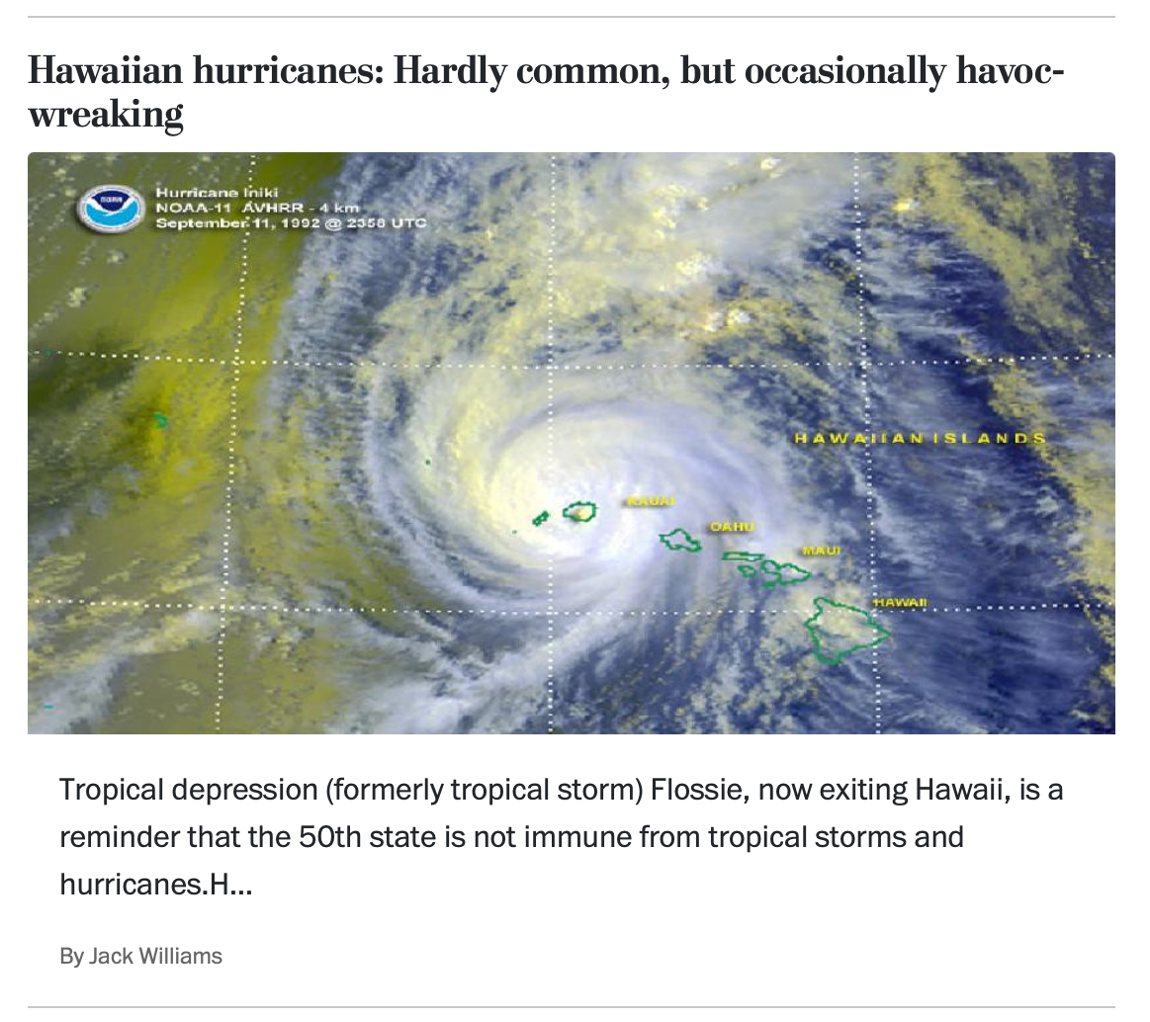}
         \caption{Title+Image+Summary (TIS): $\sim$6 rows}
         \label{fig:ct_1}
     \end{subfigure}
     \hfill
     \begin{subfigure}[b]{0.23\textwidth}
         \centering
         \includegraphics[width=\textwidth]{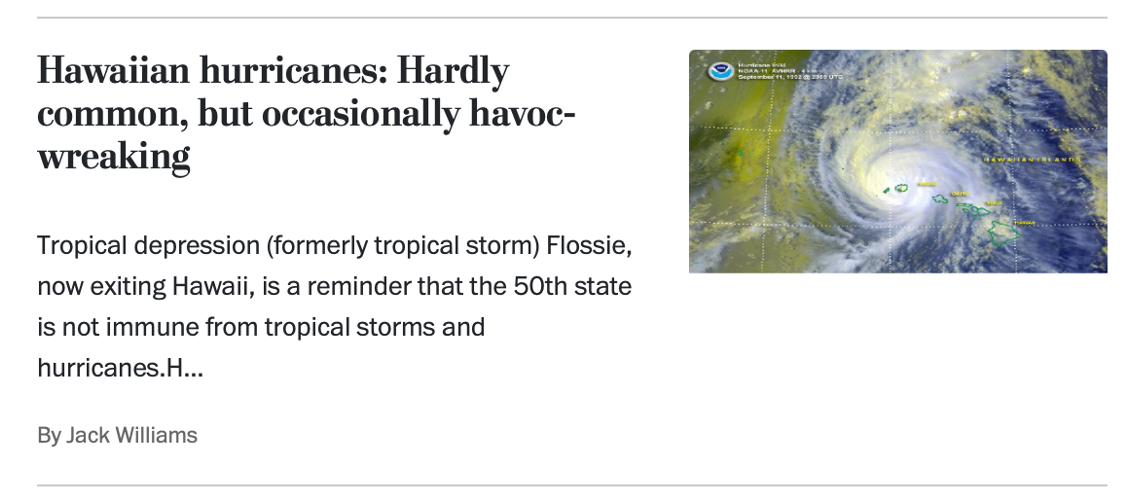}
         \caption{Title+Image+Summary (TIS WaPo): $\sim$3 rows}
         \label{fig:ct_4}
     \end{subfigure}
     \hfill
     \begin{subfigure}[b]{0.23\textwidth}
         \centering
         \includegraphics[width=\textwidth]{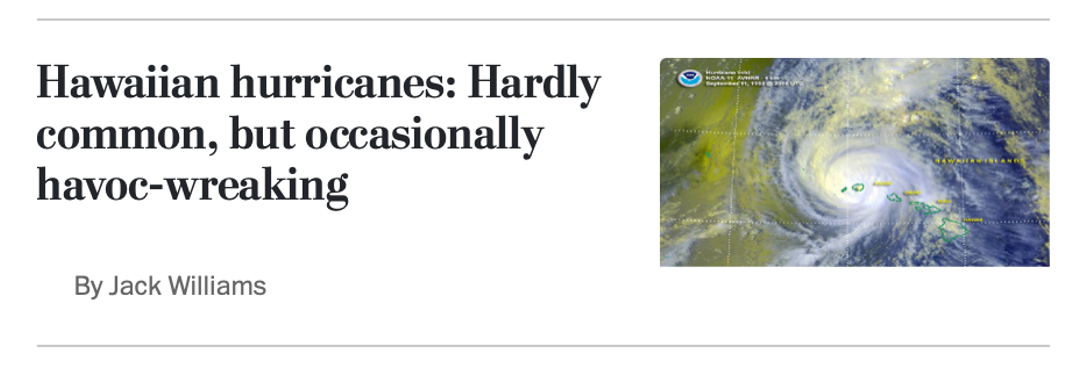}
         \caption{Title+Image (TI Google): $\sim$2 rows}
         \label{fig:ct_3}
     \end{subfigure}
     \hfill
     \begin{subfigure}[b]{0.23\textwidth}
         \centering
         \includegraphics[width=\textwidth]{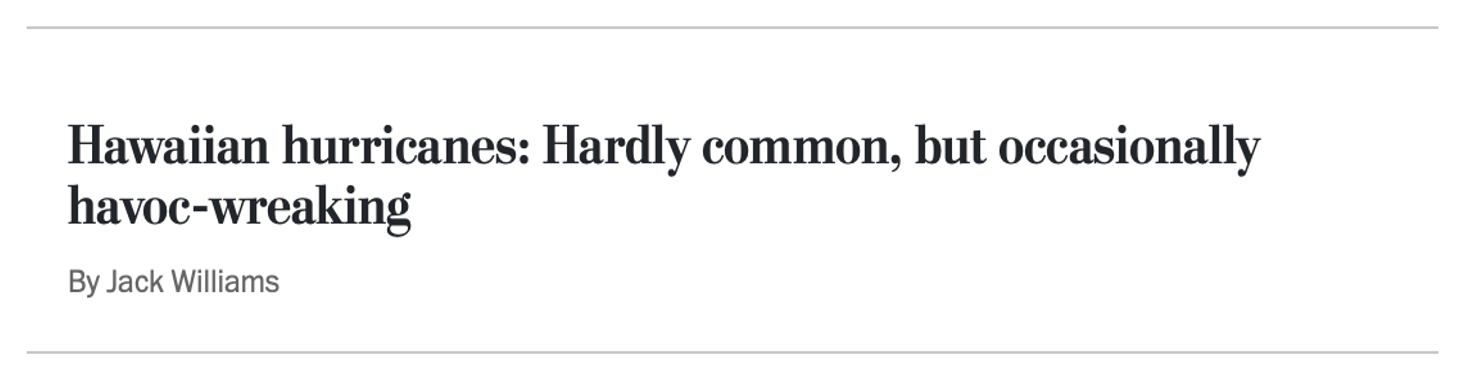}
         \caption{Title Only (T): $\sim$1 row}
         \label{fig:ct_2}
     \end{subfigure}
 \caption{Example of the different result card types, with an approximation of the number of rows each card type occupies.}
 \label{fig:card_types}
\end{figure*}

We chose five different types of interface layouts to show the participant, with the four different result cards shown in Figure~\ref{fig:card_types}.
\begin{enumerate} 
    \item Title + Image + Summary [TIS]
    \item The Washington Post Style, Title + Image + Summary [TIS WaPo],
    \item Google News Style, Title + Image[TI]
    \item Title only [T]
    \item Random, a combination of the four above.
\end{enumerate}

We consider our viewport to have a fixed amount of space (6 columns using bootstrap column widths and 12 rows, computed using approximately 100px per row). Thus, the total number of results shown on the page depended on the type of card and the number of rows it occupied. For example, on a single result page layout with our page constraint, there could either be approximately 12 T, 2 TIS cards, 6 TI Google Cards or 4 TIS WaPo cards. 

We used the same CSS stylesheet as the Washington Post website to display all the results. The titles were set at a font size of 14pt, and the summaries were set at a font size of 12pt. Since the Washington Post website summaries consisted of the leading 250 characters of the news result (at the time of the experiment), the result summaries we displayed contained the leading 250 characters as well. 

\subsection{Search Topics and Tasks}
\label{subsection:search_topics_and_tasks}
From the 50 topics available in the TREC WaPo collection, we selected four topics for our study: 
\begin{enumerate}
    \item \textbf{Topic 341:} Airport Security,
    \item \textbf{Topic 363:} Transportation Tunnel Disasters,
    \item \textbf{Topic 367:} Piracy at Sea and,
    \item \textbf{Topic 408:} Tropical Storms. 
\end{enumerate}

These topics were chosen based on their inclusion of at least 120 TREC Relevance Judgements and a minimum of 60 relevant documents with associated article images available for download. To maintain consistency in the presentation of result cards, we re-scaled each image to ensure uniform sizing. 

Participants were instructed to find and save -- different and relevant documents that they felt suited the relevance criteria for the given topic by exploring as many queries as necessary. For example, in topic 408 (see instructions in Figure~\ref{fig:serp_study_ui}(a),(b) and (c) on the left side), participants were asked to find a number of different tropical storms that caused widespread destruction and loss of life. Examples requested for the other topics were:
\begin{itemize}
    \item Topic 341 the airport and security measures employed;
    \item Topic 363 the name of the tunnel and the cause of the disaster and,
    \item Topic 367 instances of piracy where vessels were boarded.
\end{itemize}

We generated 6 queries per topic using the techniques outlined in~\cite{Maxwell2017AExperience}, and then stratified the resulting queries into three tiers based on their nDCG scores. Specifically, the queries were grouped into low (0.1-0.2), medium (0.2-0.6), and high (0.6+) nDCG categories. 

\subsection{Measures}
\label{subsection:measures}

We split the dependent variables in our study into three main categories: (a) search behaviours, (b) search experience and (c) performance: \\
\textbf{Search Behaviours: } To provide insights into user search behaviours we logged the number of...
\begin{enumerate}
    \item ...queries clicked
    \item ...pages viewed
    \item ...documents viewed
    \item ...documents saved 
\end{enumerate}

For relevant documents saved, we instructed the participant to record the relevant bits of the document into a text field that popped up if the user clicked on the ``relevant" button on the document view page and saved it to our database.
From the interaction logs, we could also compute the following time-based measures, including the time spent...
\begin{enumerate}
    \item ... to complete the task
    \item ... per result card (snippet)
    \item ... on a relevant document
    \item ... on a non-relevant document
\end{enumerate}
The relevance and non-relevance of a document were obtained using the TREC WaPo Qrels for the retrieved documents. One thing to note is that, in our document index, we only indexed documents which had TREC relevance judgements. For time spent on a snippet, we use aggregated mouse hover times as a proxy for eye gaze~\cite{Chen2001WhatBrowsing, Navalpakkam2012MouseContent, Huang2011NoSearch, Guo2010TowardsMovements, Mueller2001Cheese:Modeling} computed with a modified lightweight JavaScript code~\cite{Bhattacharya2021RecordTutorial}. 

\textbf{Search Experience: }
We measured the search experience of the participant through a user satisfaction score. We collected user satisfaction at two levels: (a) the query level (collected after changing a query) and (b) the interface level (collected after every topic/task).
For (a) query satisfaction, we collected data using a 6-point Likert scale by asking participants how satisfied they were with the results for that given query (with 1 being very dissatisfied to 6 being very satisfied). 
For (b) interface satisfaction, we asked participants whether they ...

\begin{enumerate}
\item ...felt \textbf{productive} using the system
\item ...found the interface layout to be \textbf{mentally taxing}
\item ...found the interface layout to be \textbf{engaging}
\item ...found the interface layout to be \textbf{distracting}
\item ...were \textbf{satisfied} with the interface layout,
\end{enumerate}
on a 6-point Likert scale with 1 being strongly disagree and 6 being strongly agree.

\textbf{Performance: }
By using the TREC Common Core 2018 relevance judgements, we were also able to provide an estimate of search performance at the (a) system side and (b) user side. On the system side, for each query that was submitted by a participant, we evaluated the query’s nDCG@10 and Total gain on the Page (see §\ref{rq1} for further detail). For the user-side performance measures, given all of the documents that participants clicked on and saved, we could use the aforementioned relevance judgements as ground truth, allowing us to compute the accuracy of a participant’s searching ability. This was summarised as the proportion of correctly identified relevant items saved (i.e., documents that are identified as relevant in the relevance judgements) vs. the total number saved. 

\subsection{Procedure}
\label{subsection:procedure}
Participants were recruited from the online crowd-sourcing platform Prolific \footnote{https://www.prolific.co}. Participants were also pre-screened based on their first language; all participants indicated a native speaker proficiency in English at the time of undertaking the experiment. This was done to maintain consistency across the participant's ability to carry out the task accurately. Four different pages were created on Prolific to fill participants for each topic. Each page contained the link to complete the task using the topic specified for that page. Before participating in the study, participants were presented with an on-screen information sheet detailing the procedure of the study. They were required to provide their informed consent before proceeding with the study. Upon successful completion of the study, participants received the equivalent of USD\$7 for their time, which falls in line with minimum payment requirements. Each participant was randomly allocated one of the five layouts when they began the experiment. 

The goal of the experiment was to complete a news search task based on a given topic. Participants were asked to find and mark documents relevant to one of the selected topics by exploring a set of pre-defined queries as described in §~\ref{subsection:search_topics_and_tasks}. When the participants began the experiment, they were presented with a list of six queries in a 3x2 grid that corresponded to the topic they selected. The order in which these queries were presented was randomized. An example of this query selection grid can be observed in Figure~\ref{fig:serp_study_ui}(a). Participants were instructed to choose any query to inspect and explore the associated results with that query to find the relevant documents. 
Participants were asked to evaluate the relevance of the documents based on the criteria provided on the left of the screen in a floating instruction box. An example of this floating instruction box can be seen in Figure~\ref{fig:serp_study_ui}(a),(b) and (c). These instructions were continuously visible during the process of the experiment. Once a participant picked a query, they were shown all the documents associated with that query in one of the layouts, in the style of a SERP. 

The ordering of the relevance of results was random in all layouts. In figure~\ref{fig:serp_study_ui}(b) we can see on the SERP how results were presented for a random layout, we can observe the results presented as TI, T, TIS WaPo and T. Pagination was made available via a button at the bottom of the screen to move to the next set of results for a query. The participant could click on any result card to inspect the document behind it in further detail. Upon inspecting a card, the full contents of the document were displayed on a new page, this can be seen from Figure~\ref{fig:serp_study_ui}(c). If a participant inspected a result card and found the document relevant, they were asked to provide instances of the document that made it relevant in a pop-up text area. Participants were asked to provide at least one instance per relevant document. 

When a participant moved between queries of the same topic, we collected the query satisfaction. In the query selection view, on the right side, we displayed the titles of the documents that the participant had marked as relevant, along with what section they marked within that document so that participants could quickly glance at their task progression. The participant needed to inspect at least two queries and find seven different relevant documents before finishing the topic. 
When participants finished one topic we collected the interface satisfaction as described in §~\ref{subsection:measures}, and then the second topic was randomly assigned to them (from the pool of three remaining topics) with the same result layout as the first topic.

\subsection{Participant Demographics}
Participation was completely remote, with the researchers not interacting with any participant in any capacity. Participants directly interacted with the web application designed to collect interaction data. 

The user study involved 164 participants, most of whom fell in the age range of 20 to 40 years old, with a mix of students (27) and non-students (137). The majority of participants were employed, with 122 reporting full-time or part-time work, while the remaining participants were not engaged in paid work, such as homemakers, retired, or disabled individuals.

\subsection{Ethics Approval}
A departmental review board approved the study (ethics no 2027). We strictly followed ethical guidelines and ensured that every participant gave informed consent. All participants received a thorough explanation of the study's procedures, potential risks, their rights, and the option to leave at any point. The consent form also provided a link to the ethics application approval.

\section{Experimental Results}
\label{results}

\subsection{Summary of Search Behaviours}
\begin{table*}[t]
\centering
\caption{Search behaviours, with the mean number of actions performed per user, per topic, per query. Here, \(\mathcal{Q}\) denotes Queries, \(Docs\) denotes documents. \(R\) and \(\bar{R}\) denote relevant and non-relevant. Highest accuracy values are bolded}
\begin{tabular}{l|c|c|c|c|c|c}
\toprule
\textbf{Interface Layout} & \textbf{\#$\mathcal{Q}$} & \textbf{\#$\mathbf{Docs}$ viewed} & \textbf{\#Pages} & \multicolumn{2}{c|}{\textbf{Documents Saved}} & \textbf{Accuracy} \\ 
 & & & & \textbf{\#$Docs$} & \textbf{\#$R\ Docs$} & \\ \midrule
a. TIS          & 4.12±2.19 & 4.71±4.03 & 1.05±0.22 & 2.71±2.24 & 2.19±1.69 & 0.79±0.27 \\ 
b. TIS WaPo     & 4.34±3.00 & 4.65±3.73 & 1.12±0.39 & 3.31±2.43 & 2.69±2.01 & \textbf{0.83±0.22} \\ 
c. TI Google    & 4.17±1.83 & 4.94±5.14 & 1.01±0.00 & 2.97±2.25 & 2.33±1.65 & 0.79±0.26 \\ 
d. T            & 3.91±2.33 & 5.22±4.67 & 1.05±0.24 & 2.94±2.15 & 2.40±1.74 & 0.75±0.28 \\ 
e. Random       & 4.17±2.04 & 4.63±4.31 & 1.02±0.13 & 2.89±2.96 & 2.38±2.21 & 0.78±0.25 \\ 
\bottomrule
\end{tabular}
\label{tab:search_behaviours}
\end{table*}

\begin{table*}[t]
\centering
\caption{Average timings for various search behaviours actions during the study, per user, per topic, per query. The timing data is in seconds. Asterisks (*) denote a significant difference between all groups($p < 0.05$)}
\begin{tabular}{l|c|c|c|c}
\toprule
\textbf{Interface Layout} & \textbf{Task} & \multicolumn{3}{c}{\textbf{Time per ...}} \\ 
 & & \textbf{Snippet} & $\mathbf{R\ Doc}$ & $\mathbf{\bar{R}\ Doc}$ \\ \midrule
a. TIS & $1345.23 \pm 670.13$ & $2.25 \pm 1.23^*$ & $44.13 \pm 41.90$ & $37.92 \pm 30.28$ \\ 
b. TIS WaPo & $1469.89 \pm 1138.38$ & $2.09 \pm 1.25^*$ & $52.34 \pm 45.74$ & $36.94 \pm 33.55$ \\ 
c. TI Google & $1442.04 \pm 931.42$ & $1.82 \pm 1.18^*$ & $41.24 \pm 36.48$ & $34.44 \pm 36.26$ \\ 
d. T & $1519.73 \pm 1027.33$ & $1.95 \pm 1.18^*$ & $42.36 \pm 40.77$ & $52.29 \pm 65.62$ \\ 
e. Random & $1367.29 \pm 882.38$ & $2.11 \pm 1.27^*$ & $42.97 \pm 43.16$ & $36.00 \pm 29.78$ \\ 
\bottomrule
\end{tabular}
\label{tab:timing_behaviours}
\end{table*}

Comprehensive data analysis examined differences in task completion rates, interaction times, and other user metrics, such as the number of queries, clicks, and time spent across various interface layouts. Welch's ANOVAs was used to assess whether significant differences existed between the conditions and the measures under investigation.
The primary effects were analyzed at a significance level of $\alpha = 0.05$. Pairwise Games-Howell tests were utilized for post-hoc analyses. For the reported tests, the F-score, p-value, and effect size $\eta_p^2$ are presented to two decimal places. The ranges of $\eta_p^2$ values correspond to small (< 0.06), medium (0.06 - 0.14), and large (> 0.14) effect size~\cite{Cohen1973Eta-squaredDesigns}. The $\pm$ values reported in the tables denote the mean and standard deviation.

Table \ref{tab:search_behaviours} reports the average search behaviours of users for each interface layout, detailing the number of actions performed per topic, per query. Incorporated in this analysis is the accuracy measure, highlighting how well participants identified relevant documents from the non-relevant (i.e., the proportion of relevant documents saved versus the total number saved).

Considering the varied interface layouts, there is evident consistency in user behaviours. Across the board, for any topic, participants on average clicked to view 3 to 4 queries. 
Notably, participants examined on average only a single page for every query they issued. This is despite the fact that users could examine more pages within the same query. For every query viewed, participants clicked and viewed between 4 to 5 documents. They saved about 3 of the viewed documents, and out of these, they correctly identified around 2 as relevant. The accuracy of judgements fluctuated between 0.75 to 0.83. 

We found that with the TIS WaPo layout (when all results were presented with TIS WaPo cards) participants were able to more accurately identify and mark relevant documents, achieving a peak accuracy of about 0.83, which was significantly more than other layouts (F(4,441.837) = 2.51, $p=0.04$, $\eta_p^2$ = 0.01). This is possibly due to TIS WaPo cards providing  useful information in the form of a summary that helped users to click and accurately mark them as relevant. However, it is interesting to note that this was significantly higher than the TIS layout, in which the result cards contained the same information but occupied more space. We hypothesize that this occurs due to the ability to view more cards containing summaries within the same space, potentially expanding the context window of users viewing the result cards.
However, on average, per query and topic, we found no statistically significant differences in the search behaviours of participants across any layout. 

Due to the synthetic nature of our queries and the controlled nature of our study, we hypothesize that these behaviours may be specific to our study and that in a more naturalistic search scenario, where users can type out queries, they may tend to issue queries differently to find relevant information. This could further impact other factors such as the time spent examining documents and inspecting pages on the SERP.

Table \ref{tab:timing_behaviours} offers a comprehensive look at the average timings for the search behaviours (in seconds) participants took for various actions during their search sessions. Firstly, in general, we observe that there is no statistically significant difference between the times that users took to complete the task (topic). Participants took on average approximately 20 minutes to annotate a topic. The time spent on a snippet in a layout was computed as the amount of time users spent hovering over results. We found significant differences between all layouts (F(4,9579.212) = 34.306, $p<0.001$, $\eta_p^2$ = 0.01). 
We found no notable differences in the time required to read and make a decision for a relevant or non-relevant document for any given interface layout. Participants spent an average of 43 seconds to read the document and decide the relevance.

\subsection{RQ 1: How do the quality of search results (as measured by query performance) and the interface layout impact user satisfaction in information retrieval tasks?}
\label{rq1}
\begin{figure}
    \centering
    \includegraphics[width=0.40\textwidth]{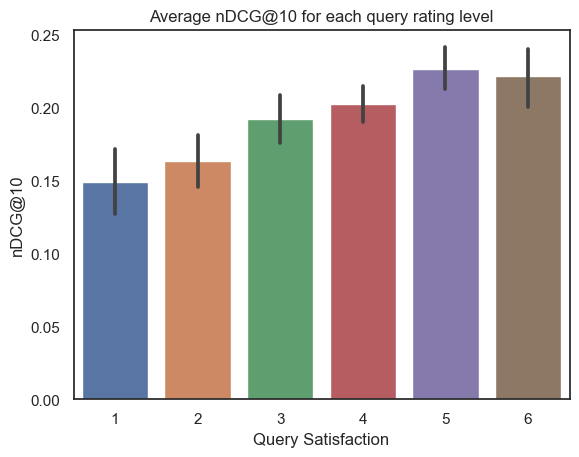}
    \caption{The relationship between query satisfaction and nDCG@10}
    \label{fig:q_sat_v_ndcg10}
\end{figure}

We ran two ordered models on 1,398 observations from 164 participants to scrutinize the association between query performance, presentation and user satisfaction. We were concerned with examining two main metrics to measure query performance. (1) nDCG@10 and (2)Total gain of the first result page.

Within our models, we also explored several interaction effects, such as the interplay between the topic and interface layout, the sequence in which the topic was completed (referred to as "topic order", meaning if the topic was completed as the first or second), and the relationship between the interface layout and query performance. Equation \ref{eqn: ndcg_10_eq} shows the independent variables in our ordered model alongside the coefficients $\beta$. For analyses focusing on the total gain on the first page, we adjusted the equation by substituting the $\beta_{1}$ coefficient.
\begin{equation}
\label{eqn: ndcg_10_eq}
\begin{aligned}
Y_{ij} &= \beta_0 + \beta_1 (\text{nDCG@10})_{ij} + \beta_2 (\text{Topic Order})_{ij} \\
&\phantom{=}+ \beta_3 (\text{Topic ID})_{ij} + \beta_4 (\text{Interface Layout})_{ij} \\
&\phantom{=}+ \beta_5 (\text{Topic ID} \times \text{Topic Order})_{ij} \\
&\phantom{=}+ \beta_6 (\text{Topic ID} \times \text{Interface Type})_{ij} \\
&\phantom{=}+ \beta_7 (\text{Interface Type} \times \text{nDCG@10})_{ij} \\
&\phantom{=}+ b_{0j} + (1 | \text{user})_{j} + \epsilon_{ij}
\end{aligned}
\end{equation}

\begin{table*}[ht]
\centering
\caption{Results of the Ordered Model analysis on query satisfaction, where p-value was statistically significant for the \(\beta\) parameter. The category differences were all significant.}
\begin{tabular}{l|c|c|c|c|c c} 
\toprule
\textbf{Beta Parameter} & \textbf{Coeff.} & \textbf{SE} & \textbf{z-value} & \textbf{p-value} & \multicolumn{2}{c}{\textbf{95\% CI}}  \\ 
\midrule
\(\beta_{1}(\text{nDCG@10})\) & 2.3015 & 0.897 & 2.566 & 0.010 & 0.543 & 4.06 \\ 
\(\beta_{5}(\text{TOPIC 408}\times\text{Order = 2})\) & 0.8355 & 0.289 & 2.891 & 0.004 & 0.269 & 1.402 \\ 
\(\beta_{6}(\text{TOPIC 408}\times\text{Interface Layout = Random})\) & 1.5771 & 0.452 & 3.490 & <0.001 & 0.691 & 2.463 \\ 

\midrule
\(\beta_{1}(\text{Total Gain on Page 1})\) & 0.2002 & 0.079 & 2.542 & 0.011 & 0.046 & 0.355 \\ 
\(\beta_{3}(\text{TOPIC 408}\times\text{Interface layout = Random})\) & 1.5491 & 0.451 & 3.434 & 0.001 & 0.665 & 2.433 \\ 
\(\beta_{5}(\text{TOPIC 408}\times\text{Order = 2})\) & 0.8471 & 0.289 & 2.936 & 0.003 & 0.282 & 1.413 \\ 
\(\beta_{7}(\text{Interface Layout = TIS WaPo}\times\text{Total Gain on Page 1})\) & 0.5173 & 0.186 & 2.778 & 0.005 & 0.152 & 0.882 \\ 
\(\beta_{7}(\text{Interface Layout = TI Google}\times\text{Total Gain on Page 1})\) & 0.3024 & 0.152 & 1.990 & 0.047 & 0.005 & 0.600 \\ 
\bottomrule
\end{tabular}
\label{tab: rq1_table}
\end{table*}

As we can observe from the top half of Table \ref{tab: rq1_table}, we found a significant positive relationship between nDCG@10 and user satisfaction, which can also be observed from Figure \ref{fig:q_sat_v_ndcg10}. We observed no interaction effects between the nDCG@10 and the interface layout which could have affected the query satisfaction. This signifies that a poorer nDCG of a query cannot increase user satisfaction to match the same level as that of a higher nDCG if we change the presentation of results. However, we observed a significant effect on query satisfaction when users attempted Topic 408 with the Random layout as the second topic. 

\begin{figure}
    \centering
    \includegraphics[width=0.45\textwidth]{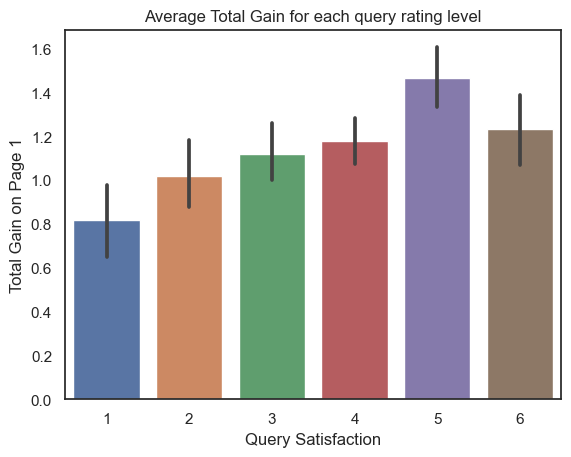}
    \caption{The relationship between query satisfaction and Total Gain on Page 1}
    \label{fig:total_gain_graph}
\end{figure}

For the total gain on the first page, we examined the effectiveness of each query within the context of the first page of results, utilizing the metric NDCG@k. Recognizing the dual significance of result relevance and quantity, we used a 'total gain' measure for the first page of results. This measure was calculated by multiplying the NDCG@k score, which evaluates the relevance of the documents on the first page, by the number of results (k) displayed on that page. By doing so, this 'total gain' measure accounts for both the quality and quantity of results, providing a more holistic assessment of query performance on the first page of results for across multiple queries. This allows us to factor in the varying number of results displayed by different interface layouts, and understand how these layouts perform not just in terms of relevance per document (as captured by NDCG@k), but also in terms of total relevance gain for the user across multiple queries. We can also observe this similarly positive relationship for the total gain from Figure~\ref{fig:total_gain_graph}

The second ordered model was run with the formula defined in Equation \ref{eqn: ndcg_10_eq}, but with the $\beta_{1}$ parameter being substituted for the total gain on page 1. The results from this second model (as shown in the bottom half of Table \ref{tab: rq1_table}) showed that the interaction effects between the total gain and the interface layouts consisting of TI Google and TIS WaPo cards played a significant effect ($p < 0.05$) on the query satisfaction. Same as with our first ordered model, we observed a significant effect on query satisfaction when users attempted topic 408 with the random layout as the second topic. Our findings from this model essentially indicate that for total gain on the first page, the presentation of results can affect user satisfaction (i.e., by modifying the interface layout, a layout with a lesser total gain on the first page can attain user satisfaction comparable to a layout with a higher total gain.)

Our results on the link between nDCG@10 and user satisfaction diverge slightly from past studies, such as \citet{Al-Maskari2007TheSatisfaction}, which found only weak ties between nDCG and satisfaction\footnote{We also compared other metrics from the \citet{Al-Maskari2007TheSatisfaction} study such as precision and CG and confirmed that precision and CG are strongly correlated to user satisfaction ($p < 0.05$) but the interface layout did not affect the user satisfaction.}.
We identified strong linear correlations between nDCG@10 and query satisfaction. Additionally, we noted distinct gains on the first page for two layouts, TI Google and TIS WaPo, revealing an interplay between result presentation, total gain, and query satisfaction. In conclusion, while nDCG@10 effectively predicts user satisfaction, no direct linear relationship exists between result presentation and user satisfaction for metrics like nDCG@10. However, metrics like total gain on the first page do influence presentation and satisfaction.

\subsection{RQ 2: What are the effects of different interface layouts on user satisfaction as measured by overall satisfaction, the likability of the engine, productivity, and mental effort?}
\begin{table*}[t]
\centering
\caption{Results of Interface Satisfaction. No statistically significant differences were found between any of the measures for a given interface layout.}
\begin{tabular}{l|c|c|c|c|c c}
\toprule
\textbf{Interface Layout} & \textbf{Felt Productive} & \textbf{Mentally Taxing} & \textbf{Liked Engine} & \textbf{Distracting} & \textbf{Overall Satisfaction} \\ 
\midrule
TIS & 3.71±1.47 & 3.52±1.51 & 3.78±1.24 & 2.95±1.31 & 3.94±1.37 \\ 
TIS WaPo & 4.05±1.58 & 3.19±1.31 & 4.00±1.22 & 2.81±1.34 & 3.92±1.35 \\ 
TI Google & 4.06±1.18 & 3.01±1.40 & 3.8±01.12 & 2.79±1.27 & 4.07±1.17 \\ 
T & 4.00±1.13 & 3.32±1.30 & 3.81±0.97 & 2.91±1.06 & 4.16±1.05 \\ 
Random & 3.78±1.33 & 3.07±1.38 & 3.85±1.13 & 2.9±1.40 & 4.10±1.17 \\ 
\bottomrule
\end{tabular}
\label{tab: interface_feedback}
\end{table*}

Looking at Table \ref{tab: interface_feedback}, we see the average satisfaction scores at the interface satisfaction for each aspect we considered. When we directly compare the layouts based on these individual metrics, the Welch ANOVA test reveals that there is no statistically significant difference between them. Since the differences might be more subtle or complex, to gain a better understanding, we used a MANOVA test.

Our analysis revealed significant differences across the different layouts. For the test statistics, including Wilks' lambda, Pillai's trace, Hotelling-Lawley trace, and Roy's greatest root, we found \( F(5, 318) = 131.647, p < 0.001 \). The observed effect sizes (\( \eta_p^2 \)) ranged from medium (0.065 for Wilks' lambda and 0.135 for Pillai's trace) to large (0.414 for both Hotelling-Lawley trace and Roy's greatest root).

Given these differences exist, we try to separate the contributing components to each interface layout via Linear Discriminant Analysis (LDA). The coefficients from the LDA, which are provided in Table \ref{tab:lda_coefficients}, represent the standardized contribution of each user satisfaction metric to the discriminant of the interface layouts.

From Table \ref{tab:lda_coefficients}, the LDA coefficients underscore that variations in interface designs subtly impacted user perceptions and experiences, culminating in different satisfaction levels, productivity perceptions, and cognitive demands. 
For instance, the T layout was predominantly associated with high overall satisfaction (0.296) and cognitive load (0.106). In contrast, the random layout interface layout was characterized by higher overall satisfaction (0.312) and lower cognitive load (-0.167), but lower productivity (-0.341), revealing a potential trade-off between user satisfaction and perceived productivity. 

\begin{figure}[h]
    \centering
    \includegraphics[width=0.45\textwidth]{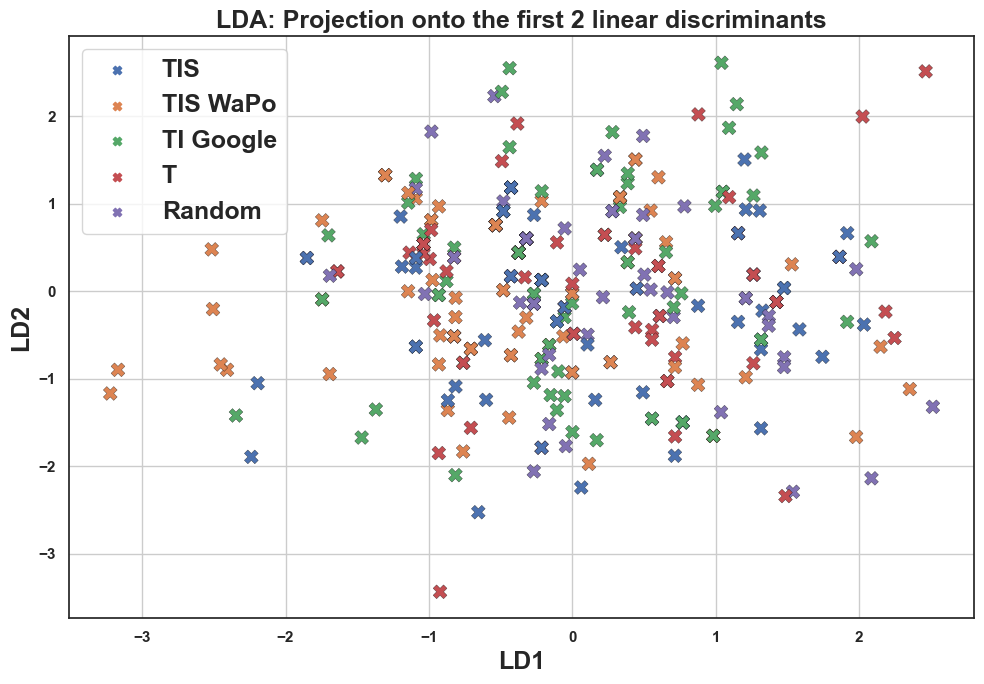}
    \caption{Visualization of the first two Linear Discriminants (LD1 and LD2), for different interface layouts}
    \label{fig:lda_plot}
\end{figure}

The explained variance ratios from the LDA show the proportion of variance captured by each discriminant function. Specifically, the first discriminant function accounts for approximately 54.8\% of the variance, highlighting its significance in distinguishing between the interface types. This is followed by the second, third, and fourth functions, which capture 26.4\%, 16.6\%, and 2.2\% of the variance, respectively. Based on these ratios of the discriminants, Figure \ref{fig:lda_plot} shows a visualization of these two discriminants in separating the different interface layouts.

\begin{table*}[t]
\centering
\caption{Coefficients of the Linear Discriminant Analysis (LDA) for distinguishing between different interface layouts based on the features captured in the interface feedback. Each row represents the coefficients for a specific interface type.}
\begin{tabular}{l|c|c|c|c|c}
\toprule
\textbf{Interface Layout} & \textbf{Felt Productive} & \textbf{Mentally Taxing} & \textbf{Liked Engine} & \textbf{Distracting} & \textbf{Overall Satisfaction} \\
\midrule
TIS & -0.180 & 0.209 & 0.030 & -0.105 & 0.133 \\
TIS WaPo & 0.260 & -0.013 & 0.349 & -0.001 & -0.559 \\
TI Google & 0.162 & -0.120 & -0.215 & -0.003 & -0.033 \\
T & 0.016 & 0.106 & -0.237 & -0.002 & 0.296 \\
Random & -0.341 & -0.167 & 0.014 & 0.119 & 0.312 \\
\bottomrule
\end{tabular}
\label{tab:lda_coefficients}
\end{table*}

Our assessment reveals that although satisfaction metrics are interconnected, they do not completely linearly differentiate the interface layouts and that there are small overlaps between the layouts (even though some clustering is observed), as seen from Figure~\ref{fig:lda_plot}. While the layouts exhibit distinct characteristics, their differences are not solely driven by individual satisfaction metrics. Instead, a collective, non-linear interaction of these metrics influences the differences observed across interface layouts. 

In our study, we have found some interesting insights. Based on the findings from RQ1, it is evident that nDCG@10 acts as a robust predictor for user satisfaction at the query level, with interface layout being influential when considering total gain on page 1. With RQ2, our exploration extends into understanding how these layouts influence user satisfaction when users complete a task (session level). While there exist differences in layout preferences and perceptions, our analysis using LDA revealed that the connection between satisfaction metrics and interface layout satisfaction is intricate and layered, deviating from a straightforward relationship. The layouts did not differ on any one specific metric of user satisfaction. These nuanced differences uncovered by LDA demonstrate that users’ satisfaction with interface layouts is multi-factorial, influenced by various combinations of satisfaction metrics. By integrating the insights from both research questions, we discern that optimizing user satisfaction in IR systems is not solely about enhancing query performance or refining the presentation of results. It requires a harmonious synchronization of both elements, considering the subtle intricacies in user preferences and satisfactions, offering a pathway to building more user-centric and adaptive Information Retrieval systems.

\section{Conclusion \& Future Work}
\label{conclusion}
In this paper, we explored the correlation between query performance, specifically marked by nDCG@10 scores, presentation and user satisfaction, with a user study consisting of 164 participants in an ad-hoc news search task. We aimed to bridge the gap between query performance, presentation and user satisfaction, venturing beyond independent studies such as \citet{Rele2005UsingInterfaces, Teevan2009VisualRevisitation, Kammerer2010HowInterface,Joho2006AWeb,Morrison2000Within-personBenefits,Verma2017SearchMobile,Al-Maskari2007TheSatisfaction} to encapsulate the nuances of presentation impact.
Our analysis revealed a strong and significant correlation between nDCG@10 scores and user satisfaction at the query level, deviating in findings from \citet{Al-Maskari2007TheSatisfaction}, where only a weak correlation was observed. However, we observed no direct relationship between the presentation (interface layouts), user satisfaction and query performance (with nDCG@10). Signifying that, while interface modifications impact user interactions and perceptions, they do not intrinsically augment the effectiveness of the queries for metrics such as nDCG@10, however, it does lead to changes with respect to other metrics such as the total gain on the first page. This means that with respect to presentation, the number of results and the space they occupy play a role in user satisfaction.
Despite the absence of a direct correlation between interface layouts and query performance, the presentation can still impact user satisfaction metrics—such as productivity, cognitive load, likability, distraction, and overall satisfaction, falling in line with all previous work such as \citet{Rele2005UsingInterfaces, Teevan2009VisualRevisitation, Kammerer2010HowInterface,Joho2006AWeb,Morrison2000Within-personBenefits,Verma2017SearchMobile} which reports that users perceive different result cards in different ways. We further assert that the differentiation in user satisfaction across interface layouts is complex, stemming from a multi-factorial combination of user satisfaction metrics. It is imperative to acknowledge that the interface’s structure holds substantial weight in shaping user satisfaction, even though it does not directly impact query performance.
This research, therefore, serves as a catalyst for a more nuanced understanding of the intricate dynamics between search performance, result presentation, and user satisfaction. It underscores the importance of interface layouts, stressing the role it plays in altering user interaction and satisfaction without directly altering search performance metrics such as the nDCG@10.
Currently, our findings are limited in their applicability to other domains and tasks due to the controlled nature of the study with limited topics, queries and interfaces. Therefore, in future endeavours, we intend to broaden the scope of our study by incorporating diverse tasks, document collections, and topics, aiming to assess the generalizability of our findings across varied contexts and scenarios. We also set the scene for further exploration into the differentiation of result card layouts based on different user satisfaction metrics. This study thus provides a foundation for further exploration into the intricate interplay between system performance, presentation, and user satisfaction.

\begin{acks}
    We want to thank the reviewers for their insightful suggestions and feedback and all the participants who took part in the study. The work reported here is funded by the DoSSIER project under the European Union’s Horizon 2020 research and innovation program, Marie Skłodowska-Curie grant agreement No. 860721
\end{acks}

\bibliographystyle{ACM-Reference-Format}

\bibliography{references}


\begin{thebibliography}{26}


\ifx \showCODEN    \undefined \def \showCODEN     #1{\unskip}     \fi
\ifx \showDOI      \undefined \def \showDOI       #1{#1}\fi
\ifx \showISBNx    \undefined \def \showISBNx     #1{\unskip}     \fi
\ifx \showISBNxiii \undefined \def \showISBNxiii  #1{\unskip}     \fi
\ifx \showISSN     \undefined \def \showISSN      #1{\unskip}     \fi
\ifx \showLCCN     \undefined \def \showLCCN      #1{\unskip}     \fi
\ifx \shownote     \undefined \def \shownote      #1{#1}          \fi
\ifx \showarticletitle \undefined \def \showarticletitle #1{#1}   \fi
\ifx \showURL      \undefined \def \showURL       {\relax}        \fi
\providecommand\bibfield[2]{#2}
\providecommand\bibinfo[2]{#2}
\providecommand\natexlab[1]{#1}
\providecommand\showeprint[2][]{arXiv:#2}

\bibitem[Al-Maskari and Sanderson(2010)]%
        {Al-Maskari2010ARetrieval}
\bibfield{author}{\bibinfo{person}{Azzah Al-Maskari} {and} \bibinfo{person}{Mark Sanderson}.} \bibinfo{year}{2010}\natexlab{}.
\newblock \showarticletitle{{A review of factors influencing user satisfaction in information retrieval}}.
\newblock \bibinfo{journal}{\emph{Journal of the American Society for Information Science and Technology}} \bibinfo{volume}{61}, \bibinfo{number}{5} (\bibinfo{date}{5} \bibinfo{year}{2010}), \bibinfo{pages}{859--868}.
\newblock
\showISSN{1532-2890}
\urldef\tempurl%
\url{https://doi.org/10.1002/ASI.21300}
\showDOI{\tempurl}


\bibitem[Al-Maskari et~al\mbox{.}(2007)]%
        {Al-Maskari2007TheSatisfaction}
\bibfield{author}{\bibinfo{person}{Azzah Al-Maskari}, \bibinfo{person}{Mark Sanderson}, {and} \bibinfo{person}{Paul Clough}.} \bibinfo{year}{2007}\natexlab{}.
\newblock \showarticletitle{{The relationship between IR effectiveness measures and user satisfaction}}.
\newblock \bibinfo{journal}{\emph{Proceedings of the 30th Annual International ACM SIGIR Conference on Research and Development in Information Retrieval, SIGIR'07}} (\bibinfo{year}{2007}), \bibinfo{pages}{773--774}.
\newblock
\showISBNx{1595935975}
\urldef\tempurl%
\url{https://doi.org/10.1145/1277741.1277902}
\showDOI{\tempurl}


\bibitem[Azzopardi(2011)]%
        {Azzopardi2011TheRetrievalb}
\bibfield{author}{\bibinfo{person}{Leif Azzopardi}.} \bibinfo{year}{2011}\natexlab{}.
\newblock \showarticletitle{{The economics in interactive information retrieval}}.
\newblock \bibinfo{journal}{\emph{SIGIR'11 - Proceedings of the 34th International ACM SIGIR Conference on Research and Development in Information Retrieval}} (\bibinfo{year}{2011}), \bibinfo{pages}{15--24}.
\newblock
\showISBNx{9781450309349}
\urldef\tempurl%
\url{https://doi.org/10.1145/2009916.2009923}
\showDOI{\tempurl}


\bibitem[Azzopardi et~al\mbox{.}(2013)]%
        {Azzopardi2013HowBehavior}
\bibfield{author}{\bibinfo{person}{Leif Azzopardi}, \bibinfo{person}{Diane Kelly}, {and} \bibinfo{person}{Kathy Brennan}.} \bibinfo{year}{2013}\natexlab{}.
\newblock \showarticletitle{{How Query Cost Affects Search Behavior}}.
\newblock \bibinfo{journal}{\emph{Proceedings of the 36th International ACM SIGIR Conference on Research and Development in Information Retrieval}} (\bibinfo{date}{7} \bibinfo{year}{2013}), \bibinfo{pages}{23--32}.
\newblock
\showISBNx{9781450320344}
\urldef\tempurl%
\url{https://doi.org/10.1145/2484028.2484049}
\showURL{%
\tempurl}


\bibitem[Azzopardi and Zuccon(2015)]%
        {Azzopardi2015AnBehavior}
\bibfield{author}{\bibinfo{person}{Leif Azzopardi} {and} \bibinfo{person}{Guido Zuccon}.} \bibinfo{year}{2015}\natexlab{}.
\newblock \showarticletitle{{An analysis of theories of search and search behavior}}. In \bibinfo{booktitle}{\emph{ICTIR 2015 - Proceedings of the 2015 ACM SIGIR International Conference on the Theory of Information Retrieval}}.
\newblock
\urldef\tempurl%
\url{https://doi.org/10.1145/2808194.2809447}
\showDOI{\tempurl}


\bibitem[Bhattacharya(2021)]%
        {Bhattacharya2021RecordTutorial}
\bibfield{author}{\bibinfo{person}{Nilavra Bhattacharya}.} \bibinfo{year}{2021}\natexlab{}.
\newblock \bibinfo{title}{{Record User Interactions on your Webpages: A tutorial}}.
\newblock
\newblock
\urldef\tempurl%
\url{https://medium.com/@nilavra/60ccc19f0516}
\showURL{%
\tempurl}


\bibitem[Borlund and Ingwersen(1997)]%
        {Borlund1997TheSystems}
\bibfield{author}{\bibinfo{person}{Pia Borlund} {and} \bibinfo{person}{Peter Ingwersen}.} \bibinfo{year}{1997}\natexlab{}.
\newblock \showarticletitle{{The development of a method for the evaluation of interactive information retrieval systems}}.
\newblock \bibinfo{journal}{\emph{Journal of Documentation}} \bibinfo{volume}{53}, \bibinfo{number}{3} (\bibinfo{year}{1997}), \bibinfo{pages}{225--250}.
\newblock
\showISSN{00220418}
\urldef\tempurl%
\url{https://doi.org/10.1108/EUM0000000007198/FULL/XML}
\showDOI{\tempurl}


\bibitem[Bota et~al\mbox{.}(2016)]%
        {Bota2016PlayingWorkload}
\bibfield{author}{\bibinfo{person}{Horaţiu Bota}, \bibinfo{person}{Ke Zhou}, {and} \bibinfo{person}{Joemon~M. Jose}.} \bibinfo{year}{2016}\natexlab{}.
\newblock \showarticletitle{{Playing your cards right: The effect of entity cards on search behaviour and workload}}.
\newblock \bibinfo{journal}{\emph{CHIIR 2016 - Proceedings of the 2016 ACM Conference on Human Information Interaction and Retrieval}} (\bibinfo{date}{3} \bibinfo{year}{2016}), \bibinfo{pages}{131--140}.
\newblock
\showISBNx{9781450337519}
\urldef\tempurl%
\url{https://doi.org/10.1145/2854946.2854967}
\showDOI{\tempurl}


\bibitem[Chen et~al\mbox{.}(2001)]%
        {Chen2001WhatBrowsing}
\bibfield{author}{\bibinfo{person}{Mon~Chu Chen}, \bibinfo{person}{John~R. Anderson}, {and} \bibinfo{person}{Myeong~Ho Sohn}.} \bibinfo{year}{2001}\natexlab{}.
\newblock \showarticletitle{{What can a mouse cursor tell us more? Correlation of eye/mouse movements on web browsing}}.
\newblock \bibinfo{journal}{\emph{Conference on Human Factors in Computing Systems - Proceedings}} (\bibinfo{year}{2001}), \bibinfo{pages}{281--282}.
\newblock
\showISBNx{1581133405}
\urldef\tempurl%
\url{https://doi.org/10.1145/634067.634234}
\showDOI{\tempurl}


\bibitem[Cohen(1973)]%
        {Cohen1973Eta-squaredDesigns}
\bibfield{author}{\bibinfo{person}{Jacob Cohen}.} \bibinfo{year}{1973}\natexlab{}.
\newblock \showarticletitle{{Eta-squared and partial eta-squared in fixed factor anova designs}}.
\newblock \bibinfo{journal}{\emph{Educational and Psychological Measurement}} \bibinfo{volume}{33}, \bibinfo{number}{1} (\bibinfo{date}{4} \bibinfo{year}{1973}), \bibinfo{pages}{107--112}.
\newblock
\showISSN{15523888}
\urldef\tempurl%
\url{https://doi.org/10.1177/001316447303300111/ASSET/001316447303300111.FP.PNG{\_}V03}
\showDOI{\tempurl}


\bibitem[Dziadosz and Chandrasekar(2002)]%
        {Dziadosz2002DoResults}
\bibfield{author}{\bibinfo{person}{Susan Dziadosz} {and} \bibinfo{person}{Raman Chandrasekar}.} \bibinfo{year}{2002}\natexlab{}.
\newblock \showarticletitle{{Do thumbnail previews help users make better relevance decisions about web search results?}}. In \bibinfo{booktitle}{\emph{SIGIR Forum (ACM Special Interest Group on Information Retrieval)}}.
\newblock
\showISSN{01635840}
\urldef\tempurl%
\url{https://doi.org/10.1145/564437.564446}
\showDOI{\tempurl}


\bibitem[Guo and Agichtein(2010)]%
        {Guo2010TowardsMovements}
\bibfield{author}{\bibinfo{person}{Qi Guo} {and} \bibinfo{person}{Eugene Agichtein}.} \bibinfo{year}{2010}\natexlab{}.
\newblock \showarticletitle{{Towards predicting web searcher gaze position from mouse movements}}.
\newblock \bibinfo{journal}{\emph{Conference on Human Factors in Computing Systems - Proceedings}} (\bibinfo{year}{2010}), \bibinfo{pages}{3601--3606}.
\newblock
\showISBNx{9781605589312}
\urldef\tempurl%
\url{https://doi.org/10.1145/1753846.1754025}
\showDOI{\tempurl}


\bibitem[Huang et~al\mbox{.}(2011)]%
        {Huang2011NoSearch}
\bibfield{author}{\bibinfo{person}{Jeff Huang}, \bibinfo{person}{Ryen~W. White}, {and} \bibinfo{person}{Susan Dumais}.} \bibinfo{year}{2011}\natexlab{}.
\newblock \showarticletitle{{No clicks, no problem: Using cursor movements to understand and improve search}}.
\newblock \bibinfo{journal}{\emph{Conference on Human Factors in Computing Systems - Proceedings}} (\bibinfo{year}{2011}), \bibinfo{pages}{1225--1234}.
\newblock
\showISBNx{9781450302289}
\urldef\tempurl%
\url{https://doi.org/10.1145/1978942.1979125}
\showDOI{\tempurl}


\bibitem[Jansen et~al\mbox{.}(2000)]%
        {Jansen2000RealWeb}
\bibfield{author}{\bibinfo{person}{Bernard~J. Jansen}, \bibinfo{person}{Amanda Spink}, {and} \bibinfo{person}{Tefko Saracevic}.} \bibinfo{year}{2000}\natexlab{}.
\newblock \showarticletitle{{Real life, real users, and real needs: A study and analysis of user queries on the Web}}.
\newblock \bibinfo{journal}{\emph{Information Processing and Management}} \bibinfo{volume}{36}, \bibinfo{number}{2} (\bibinfo{date}{3} \bibinfo{year}{2000}), \bibinfo{pages}{207--227}.
\newblock
\showISSN{03064573}
\urldef\tempurl%
\url{https://doi.org/10.1016/S0306-4573(99)00056-4}
\showDOI{\tempurl}


\bibitem[Joho and Jose(2006)]%
        {Joho2006AWeb}
\bibfield{author}{\bibinfo{person}{Hideo Joho} {and} \bibinfo{person}{Joemon~M Jose}.} \bibinfo{year}{2006}\natexlab{}.
\newblock \showarticletitle{{A comparative study of the effectiveness of search result presentation on the Web}}. In \bibinfo{booktitle}{\emph{Lecture Notes in Computer Science (including subseries Lecture Notes in Artificial Intelligence and Lecture Notes in Bioinformatics)}}, Vol.~\bibinfo{volume}{3936 LNCS}.
\newblock
\showISSN{16113349}
\urldef\tempurl%
\url{https://doi.org/10.1007/11735106{\_}27}
\showDOI{\tempurl}


\bibitem[Kammerer and Gerjets(2010)]%
        {Kammerer2010HowInterface}
\bibfield{author}{\bibinfo{person}{Yvonne Kammerer} {and} \bibinfo{person}{Peter Gerjets}.} \bibinfo{year}{2010}\natexlab{}.
\newblock \showarticletitle{{How the interface design influences users' spontaneous trustworthiness evaluations of web search results: Comparing a list and a grid interface}}. In \bibinfo{booktitle}{\emph{Eye Tracking Research and Applications Symposium (ETRA)}}.
\newblock
\urldef\tempurl%
\url{https://doi.org/10.1145/1743666.1743736}
\showDOI{\tempurl}


\bibitem[Kelly and Azzopardi(2015)]%
        {Kelly2015HowExperience}
\bibfield{author}{\bibinfo{person}{Diane Kelly} {and} \bibinfo{person}{Leif Azzopardi}.} \bibinfo{year}{2015}\natexlab{}.
\newblock \showarticletitle{{How many results per page? A study of SERP size, search behavior and user experience}}.
\newblock \bibinfo{journal}{\emph{SIGIR 2015 - Proceedings of the 38th International ACM SIGIR Conference on Research and Development in Information Retrieval}} (\bibinfo{date}{8} \bibinfo{year}{2015}), \bibinfo{pages}{183--192}.
\newblock
\showISBNx{9781450336215}
\urldef\tempurl%
\url{https://doi.org/10.1145/2766462.2767732}
\showDOI{\tempurl}


\bibitem[Maxwell et~al\mbox{.}(2017)]%
        {Maxwell2017AExperience}
\bibfield{author}{\bibinfo{person}{David Maxwell}, \bibinfo{person}{Leif Azzopardi}, {and} \bibinfo{person}{Yashar Moshfeghi}.} \bibinfo{year}{2017}\natexlab{}.
\newblock \showarticletitle{{A study of snippet length and informativeness behaviour, performance and user experience}}.
\newblock \bibinfo{journal}{\emph{SIGIR 2017 - Proceedings of the 40th International ACM SIGIR Conference on Research and Development in Information Retrieval}} (\bibinfo{date}{8} \bibinfo{year}{2017}), \bibinfo{pages}{135--144}.
\newblock
\showISBNx{9781450350228}
\urldef\tempurl%
\url{https://doi.org/10.1145/3077136.3080824}
\showDOI{\tempurl}


\bibitem[Morrison and Vancouver(2000)]%
        {Morrison2000Within-personBenefits}
\bibfield{author}{\bibinfo{person}{Elizabeth~W Morrison} {and} \bibinfo{person}{Jeffrey~B Vancouver}.} \bibinfo{year}{2000}\natexlab{}.
\newblock \showarticletitle{{Within-person analysis of information seeking: The effects of perceived costs and benefits}}.
\newblock \bibinfo{journal}{\emph{Journal of Management}} \bibinfo{volume}{26}, \bibinfo{number}{1} (\bibinfo{year}{2000}), \bibinfo{pages}{119--137}.
\newblock
\showISSN{01492063}
\urldef\tempurl%
\url{https://doi.org/10.1177/014920630002600101}
\showDOI{\tempurl}


\bibitem[Mueller and Lockerd(2001)]%
        {Mueller2001Cheese:Modeling}
\bibfield{author}{\bibinfo{person}{Florian Mueller} {and} \bibinfo{person}{Andrea Lockerd}.} \bibinfo{year}{2001}\natexlab{}.
\newblock \showarticletitle{{Cheese: Tracking mouse movement activity on websites, a tool for user modeling}}.
\newblock \bibinfo{journal}{\emph{Conference on Human Factors in Computing Systems - Proceedings}} (\bibinfo{year}{2001}), \bibinfo{pages}{279--280}.
\newblock
\showISBNx{1581133405}
\urldef\tempurl%
\url{https://doi.org/10.1145/634067.634233}
\showDOI{\tempurl}


\bibitem[Navalpakkam and Churchill(2012)]%
        {Navalpakkam2012MouseContent}
\bibfield{author}{\bibinfo{person}{Vidhya Navalpakkam} {and} \bibinfo{person}{Elizabeth~F Churchill}.} \bibinfo{year}{2012}\natexlab{}.
\newblock \showarticletitle{{Mouse Tracking: Measuring and Predicting Users' Experience of Web-based Content}}.
\newblock  (\bibinfo{year}{2012}).
\newblock
\showISBNx{9781450310154}


\bibitem[Rele and Duchowski(2005)]%
        {Rele2005UsingInterfaces}
\bibfield{author}{\bibinfo{person}{Rachana~S Rele} {and} \bibinfo{person}{Andrew~T Duchowski}.} \bibinfo{year}{2005}\natexlab{}.
\newblock \showarticletitle{{Using eye tracking to evaluate alternative search results interfaces}}. In \bibinfo{booktitle}{\emph{Proceedings of the Human Factors and Ergonomics Society}}.
\newblock
\showISSN{10711813}
\urldef\tempurl%
\url{https://doi.org/10.1177/154193120504901508}
\showDOI{\tempurl}


\bibitem[Roy et~al\mbox{.}(2022)]%
        {Roy2022UsersExperiences}
\bibfield{author}{\bibinfo{person}{Nirmal Roy}, \bibinfo{person}{David Maxwell}, {and} \bibinfo{person}{Claudia Hauff}.} \bibinfo{year}{2022}\natexlab{}.
\newblock \showarticletitle{{Users and Contemporary SERPs: A (Re-)Investigation Examining User Interactions and Experiences}}.
\newblock \bibinfo{journal}{\emph{SIGIR 2022 - Proceedings of the 45th International ACM SIGIR Conference on Research and Development in Information Retrieval}} \bibinfo{volume}{11}, \bibinfo{number}{22} (\bibinfo{date}{7} \bibinfo{year}{2022}), \bibinfo{pages}{2765--2775}.
\newblock
\urldef\tempurl%
\url{https://doi.org/10.1145/3477495.3531719}
\showDOI{\tempurl}


\bibitem[Teevan et~al\mbox{.}(2009)]%
        {Teevan2009VisualRevisitation}
\bibfield{author}{\bibinfo{person}{Jaime Teevan}, \bibinfo{person}{Edward Cutrell}, \bibinfo{person}{Danyel Fisher}, \bibinfo{person}{Steven~M Drucker}, \bibinfo{person}{Gonzalo Ramos}, \bibinfo{person}{Paul Andr{\'{e}}}, {and} \bibinfo{person}{Chang Hu}.} \bibinfo{year}{2009}\natexlab{}.
\newblock \showarticletitle{{Visual snippets: Summarizing web pages for search and revisitation}}. In \bibinfo{booktitle}{\emph{Conference on Human Factors in Computing Systems - Proceedings}}.
\newblock
\urldef\tempurl%
\url{https://doi.org/10.1145/1518701.1519008}
\showDOI{\tempurl}


\bibitem[Tombros and Sanderson(1998)]%
        {Tombros1998AdvantagesRetrieval}
\bibfield{author}{\bibinfo{person}{Anastasios Tombros} {and} \bibinfo{person}{Mark Sanderson}.} \bibinfo{year}{1998}\natexlab{}.
\newblock \showarticletitle{{Advantages of query biased summaries in information retrieval}}.
\newblock \bibinfo{journal}{\emph{SIGIR Forum (ACM Special Interest Group on Information Retrieval)}} (\bibinfo{year}{1998}).
\newblock
\showISSN{01635840}
\urldef\tempurl%
\url{https://doi.org/10.1145/290941.290947}
\showDOI{\tempurl}


\bibitem[Verma and Yilmaz(2017)]%
        {Verma2017SearchMobile}
\bibfield{author}{\bibinfo{person}{Manisha Verma} {and} \bibinfo{person}{Emine Yilmaz}.} \bibinfo{year}{2017}\natexlab{}.
\newblock \showarticletitle{{Search costs vs. User satisfaction on mobile}}.
\newblock \bibinfo{journal}{\emph{Lecture Notes in Computer Science (including subseries Lecture Notes in Artificial Intelligence and Lecture Notes in Bioinformatics)}}  \bibinfo{volume}{10193 LNCS} (\bibinfo{year}{2017}), \bibinfo{pages}{698--704}.
\newblock
\showISBNx{9783319566078}
\showISSN{16113349}
\urldef\tempurl%
\url{https://doi.org/10.1007/978-3-319-56608-5{\_}68/FIGURES/8}
\showDOI{\tempurl}


\end{thebibliography}

\end{document}